\documentclass[conference]{IEEEtran}
\ifCLASSINFOpdf
  \usepackage[pdftex]{graphicx}
  \graphicspath{{pics/}}
   \DeclareGraphicsExtensions{.tex,.pdf,.jpeg,.jpg,.png}
\else
\fi
\usepackage{tikz}

\begin{document}
%
\title{Experiments to increase the used Energy with the PEGASUS Railgun}

\author{\IEEEauthorblockN{S.\,Hundertmark\IEEEmembership{Member,~IEEE}, 
M.\,Schneider, D.\,Simicic, G.\,Vincent}
\IEEEauthorblockA{French-German Research Institute of Saint-Louis\\
68301 Saint Louis, France\\
Email: stephan.hundertmark@isl.eu}
}

%


\maketitle

\begin{abstract}
The French-German Research Institute (ISL) has several railguns
installed, the largest of these is the {\sc PEGASUS} accelerator. 
It is a 6\,m long, 4\,x\,4\,cm$^2$ caliber distributed energy supply 
(DES) railgun. 
It has a 10\,MJ capacitor bank as energy supply attached to it. In the past,
this installation was used to accelerate projectiles with a mass 
of about 300\,g to velocities up to 2500\,m/s. In the ongoing
investigation, it is attempted to accelerate heavier projectiles to 
velocities above 2000\,m/s. For this a new type of projectile including 
a payload section was developed. In this paper the results of the 
experiments with payload projectiles using a primary energy between 
3.8\,MJ and 4.8\,MJ are discussed.
\end{abstract}

\section{Introduction}
In the military domain, there are two main applications for a railgun.
Both make use of the, compared to explosively driven guns, superior
muzzle velocity being achievable. A long range electromagnetic
artillery system can cover ranges up to 400\,km. Such a system would 
accelerate a heavy projectile with a mass of up to 60\,kg to a muzzle
velocity of 2500\,m/s \cite{mcnab_1}. Alternatively one can use the
railgun to accelerate lighter projectiles in rapid succession to bring 
down approaching missiles in a last line of defense scenario. Both 
scenarios have the common requirement of a
large muzzle velocity, but very different demands concerning the
needed power levels and the design of the railgun itself. 

At the ISL, {\sc PEGASUS } is the largest railgun installation. 
This railgun has a 
primary energy storage of 10\,MJ and an overall efficiency of about 30\,\%. 
With the aim of reaching the desired 2500\,m/s,
the maximal total weight of the launch package has to stay below
1\,kg. Starting from this, different payload projectiles in the mass
range of 600\,g to 700\,g were developed and tested. In a first series
of investigations, a monolithic payload projectile was launched with
primary energies ranging from 2.6\,MJ to 3.7\,MJ, to velocities of up
to 1560\,m/s. To make better use of the primary energy and to
investigate the innovative concept of a separating projectile, a
projectile being split into two parts was designed. While the
armature of the monolithic projectile consists out 4 rows of brushes
in the rear half of the projectile, the separating projectile can
leave the part of the projectile body with the 3 rearmost brush rows
behind. At the same energy level of 3.8\,MJ the payload reached a
velocity of 1825\,m/s. A more detailed description of these experiments 
can be found in \cite{hundertmark_1}. Here further modifications of
the separating projectile and the experimental results with this type
of projectile are described. 
\section{{\sc PEGASUS} Railgun}
\begin{figure}[tb!]
\centering
\includegraphics[width=3.5in]{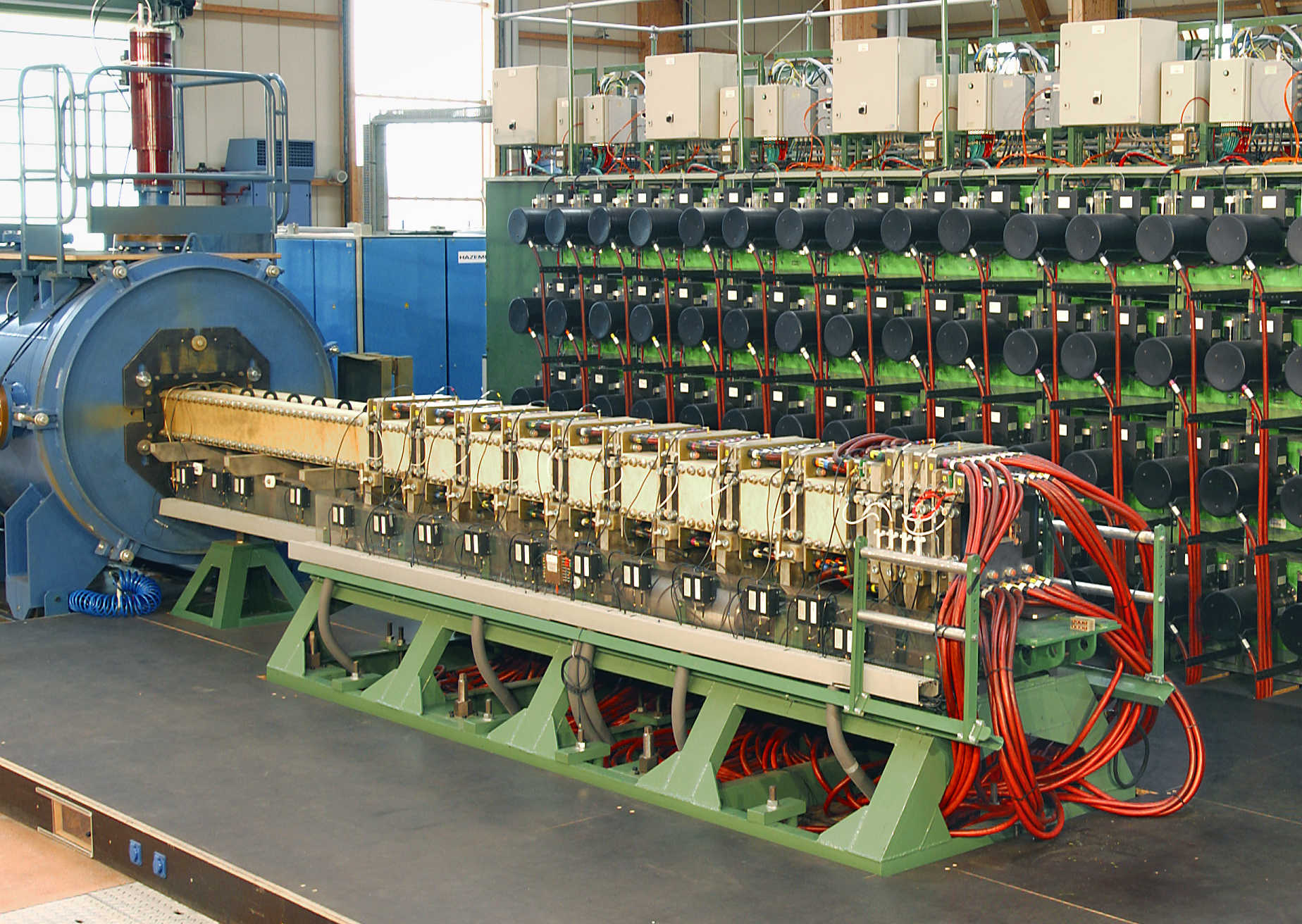}
\caption{{\sc Pegasus} railgun facility.}
\label{pegasus_1}
\end{figure}
The {\sc PEGASUS} railgun was built in 1998 with the goal to
accelerate projectiles with masses of about 1\,kg to velocities above 
2000\,m/s in the medium caliber range (40\,mm to 50\,mm). Initially it
was equipped with a glass-fiber and carbon-fiber wound barrel with a 
caliber of 50\,mm \cite{lehmann_1}. 
Figure \ref{pegasus_1} shows the {\sc PEGASUS} installation with the
currently mounted barrel, being in use since 2002 \cite{lehmann_2}. 
This barrel is 6\,m long and has a square caliber of 40\,mm. A total of up
to 10\,MJ of electrical energy can be stored in the 200 capacitor
modules being visible in the background of the railgun. 
{\sc PEGASUS} is a distributed energy supply
(DES) system connected to 13 banks with 16 (the last one only 8)
capacitor modules, each. Banks no.\,1
and no.\,2 are connected to the breech, the other banks are connected to
current injection points being distributed along the first 3.6\,m of
the barrel length. For the purpose of the experiments being performed 
for this investigation, 
4 capacitor modules in the banks no.\,8 to no.\,12 are disconnected.
This reduces the capacity of these banks by 25\,\% and the total
capacity by 10\,\%.  Each capacitor module
can store up to 50\,kJ and deliver a maximum current of 50\,kA. 
They are equipped with a high-voltage capacitor, a thyristor, a crowbar diode 
and a pulse forming coil \cite{spahn_1}. The electrical
connection of the capacitor module to the accelerator is made with a coaxial cable. 
Due to the DES scheme the maximum
total current that can be delivered to the gun is approx. 2\,MA. 
The cylinder on the left hand side of figure \ref{pegasus_1} is a 7\,m
long catch tank, being equipped with several flash x-ray tubes,
Doppler radar devices and the possibility to mount several high-speed
cameras. At the end of the free-flight phase in the cylinder, the
projectile is stopped using steel plates.

\section{{\sc Pegasus} Standard Projectile}
\begin{figure}[t]
\centering
\includegraphics[width=3.in]{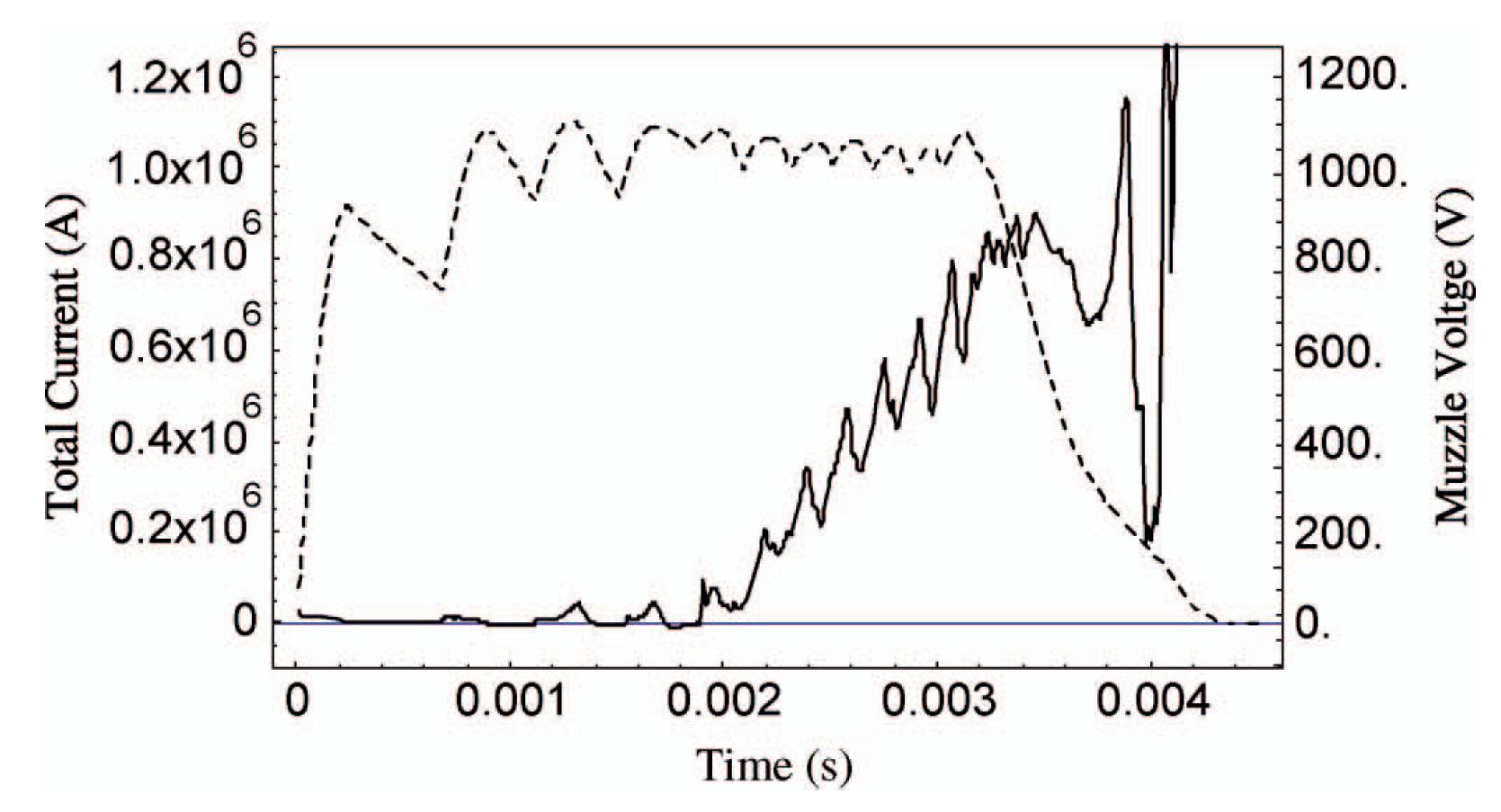}\\
\caption{Current (dashed line) an muzzle voltage (straight line) distributions for a high-velocity
(v $>$ 2500 m/s) standard projectile shot (from \cite{reck_2}).}
\label{t194_1}
\end{figure}
Since several years, projectiles used with the {\sc PEGASUS} railgun are
made out of glass fiber reinforced plastic (GRP) equipped with
copper-fiber brush armatures. The GRP body of the projectile is
lightweight, electrical isolating and relatively easy to manufacture.
In combination with the copper brush armature, the current is confined
to well defined paths through the projectile. 
In combination with appropriate sensors, this allows for
the investigation of the current distribution in between 
different rows of brushes during launch conditions \cite{liebfried_1,schneider_1}.
The standard type of projectile, that was used in many investigations
with {\sc PEGASUS}, consists out of 8 brushes arranged in 4 rows. The
total mass for this type of projectile is approx. 300\,g. The body of these projectiles
is fabricated using industrial pre impregnated E-glass laminate. With
these projectiles it was possible to reach velocities of about
2200\,m/s, for higher velocities delamination and break-up of the GRP
was observed \cite{christoph_1}. One possibility to increase the
velocity potential of plastic projectile bodies is to improve 
the stability against hard acceleration by using innovative
materials. Experiments using a body composed out of 
alternating layers of GRP and carbon fiber materials allowed to reach
a velocity well above 2500\,m/s \cite{reck_2}. In figure \ref{t194_1}
the current and muzzle voltage trace for this shot is shown. The DES system
of {\sc PEGASUS} allows to generate a rather flat current pulse. Here, the 
current driving the projectile is just below 1.1\,MA. The muzzle
voltage trace shows that
excellent sliding conditions exists until approx. 2\,ms. At that time
the voltage increases rapidly, indicating the onset of transition.
The peaks in the muzzle voltage signal, clearly seen in the raising part
of the graph (between 2\,ms and 3.5\,ms), are explained by
inductive effects connected to the current injection locations 
of the DES system of the launcher \cite{reck_1}.
Usually in experiment with {\sc PEGASUS}, the plasma after transition 
appears in between one rail and the armature, only, while the other side of 
the projectile is pressed heavily against the opposite rail.
As long as the plasma is connected to the brushes and confined in the
small space between projectile surface and rails, there is still a
well defined current path and the clear pattern of inductive peaks as
seen in the figure is visible. Underlying the peaks, there is a nearly
linear rise of the muzzle voltage. This effect can be explained by an 
increase in resistance seen by the current flowing through the plasma
and the copper brushes. Contributing to this resistance is the increase 
of the resistivity of the copper brushes due to the increase of the
temperature and possibly a dilution effect of the plasma due to the
rapid movement of the projectile.
\section{Payload Projectiles}
\begin{figure}[tb!]
\centering
\includegraphics[height=1.05in]{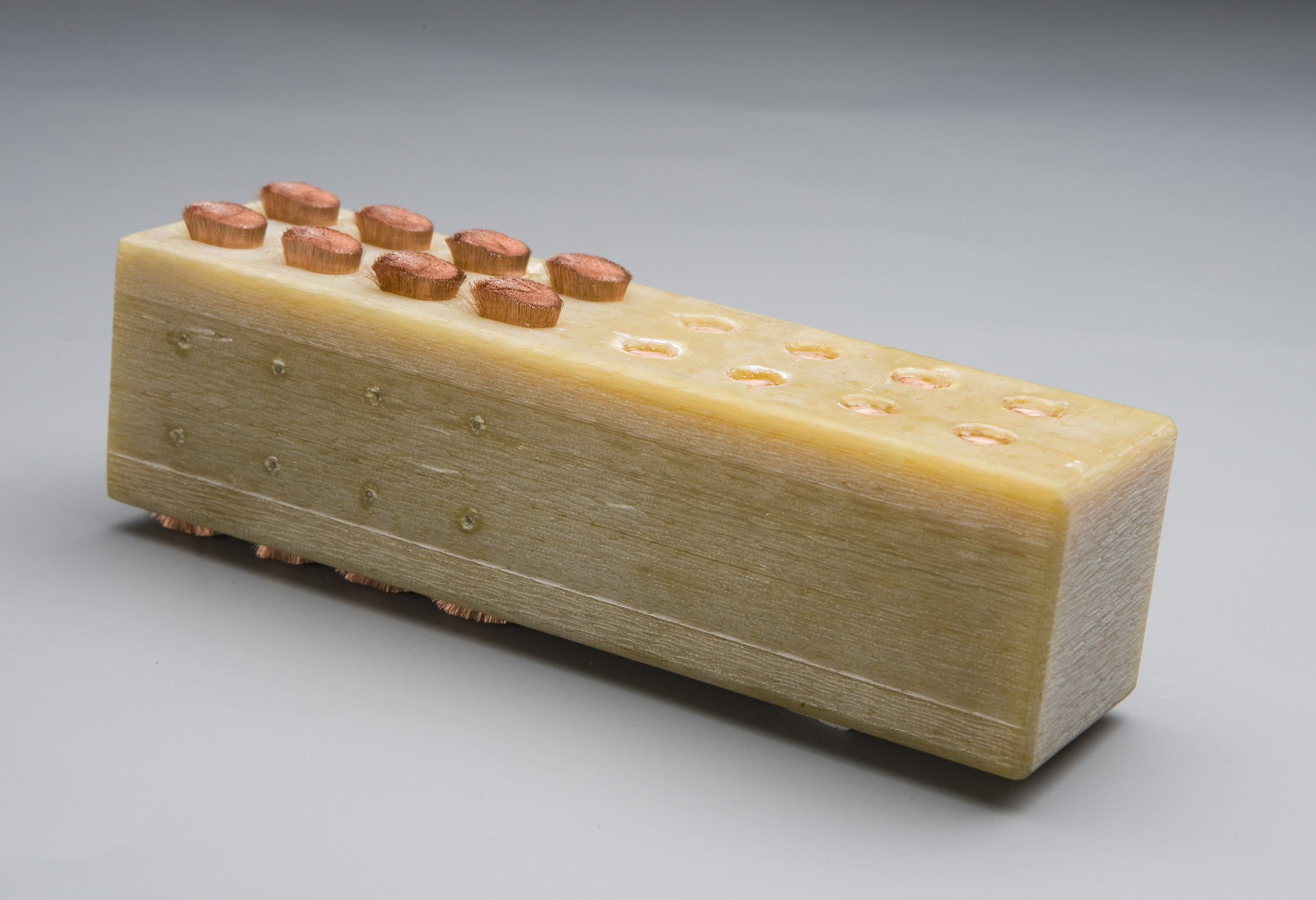}\hfill
\includegraphics[height=1.05in]{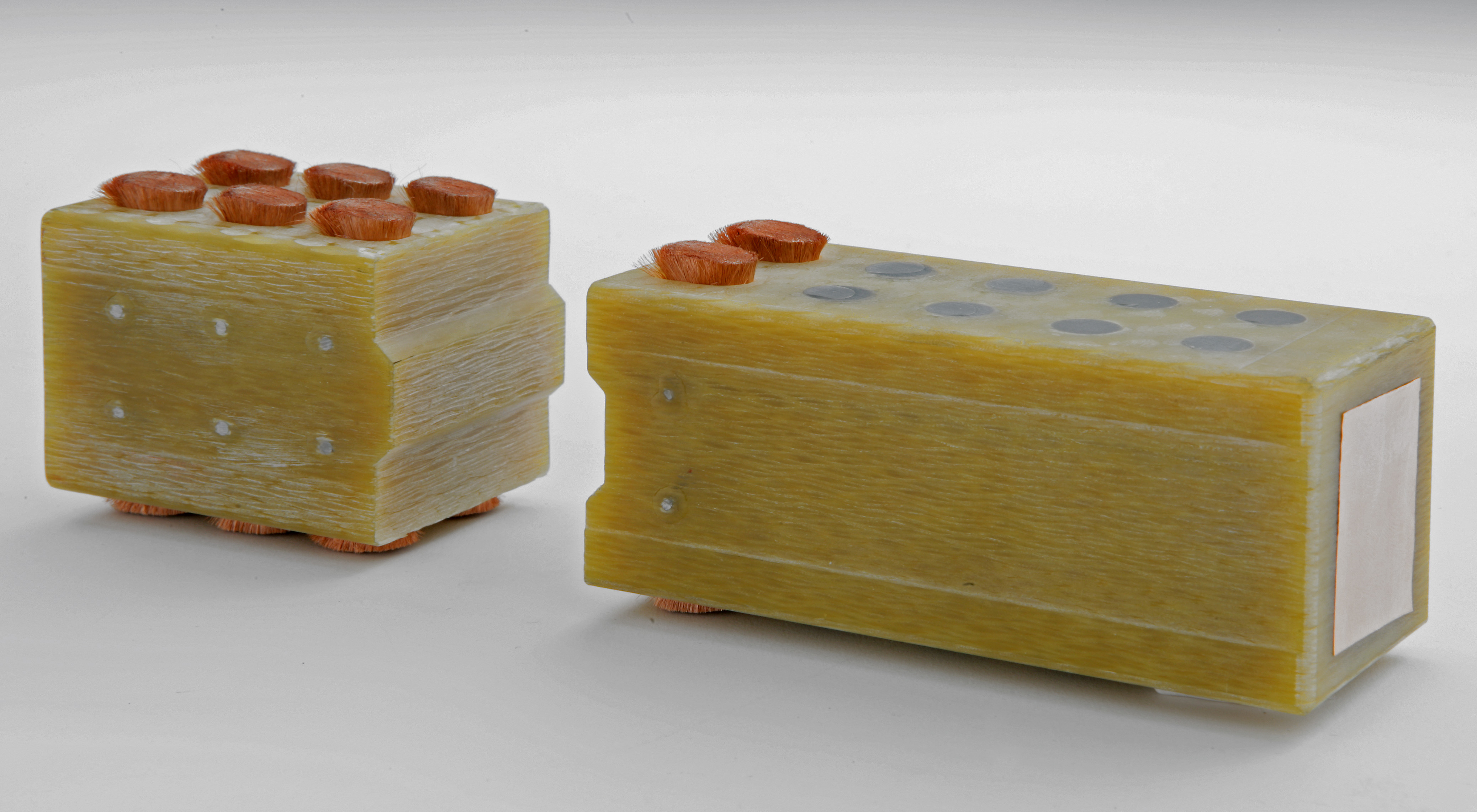}
\caption{Two versions of payload projectiles for the {\sc PEGASUS}
railgun.}
\label{pegasus_payload_projectiles}
\end{figure}
In figure \ref{pegasus_payload_projectiles} two different types of
payload projectiles are shown. The projectile on the left hand side has
a monolithic design, with a total mass of about 625\,g. The rear half
of the 140\,mm long projectile is identical to the standard projectile
and carries 8 copper brushes in 4 rows. The forward part is equipped
with 8 copper
cylinders as payload. To reach higher velocities at the same electrical 
energy distributed to the launcher, the separating projectile seen on
the right of figure \ref{pegasus_payload_projectiles} was introduced.
The basic idea behind this type of projectile is, that during the
acceleration the brushes are
eroded row-wise from the rear to the front. Once the 3 rearmost
brushes lost electrical contact to the rails, this part is left behind and only the 
payload section continues to be accelerated. The total mass and the dimensions 
for this projectile are the same as for the monolithic design.
The forward payload part with the 2 booster brushes weighs approx. 400\,g.
In \cite{hundertmark_1} experiments with the monolithic projectile are
described. A muzzle velocity of 1560\,m/s was reached using 3.8\,MJ of
electrical energy. Accelerating a separating projectile using the same
electrical energy resulted in a velocity of the payload section of 1825\,m/s,
an increase of 16\% compared to the monolithic projectile design. 
\section{Experiments with Separating Projectiles}
\begin{figure}[tb!]
\centering
\includegraphics[width=1.75in]{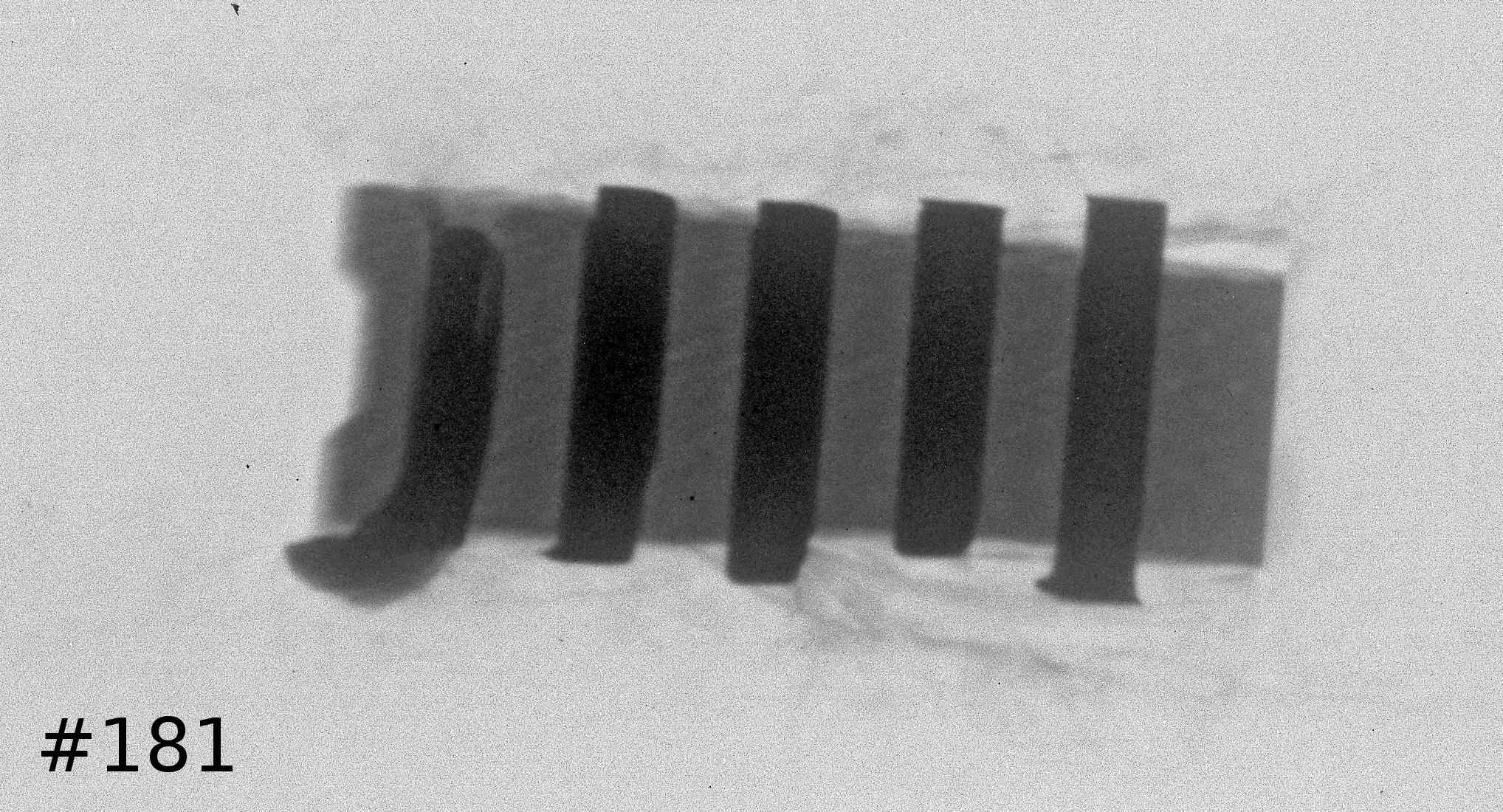}
\hfill
\includegraphics[width=1.6in]{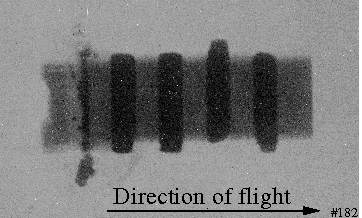}
\caption{X-ray picture of the payload part for shot no.\,181 (left)
and no.\,182 (right).}
\label{t181_t182_xray}
\end{figure}

\begin{figure}[tb!]
\centering
\includegraphics[width=.48\textwidth]{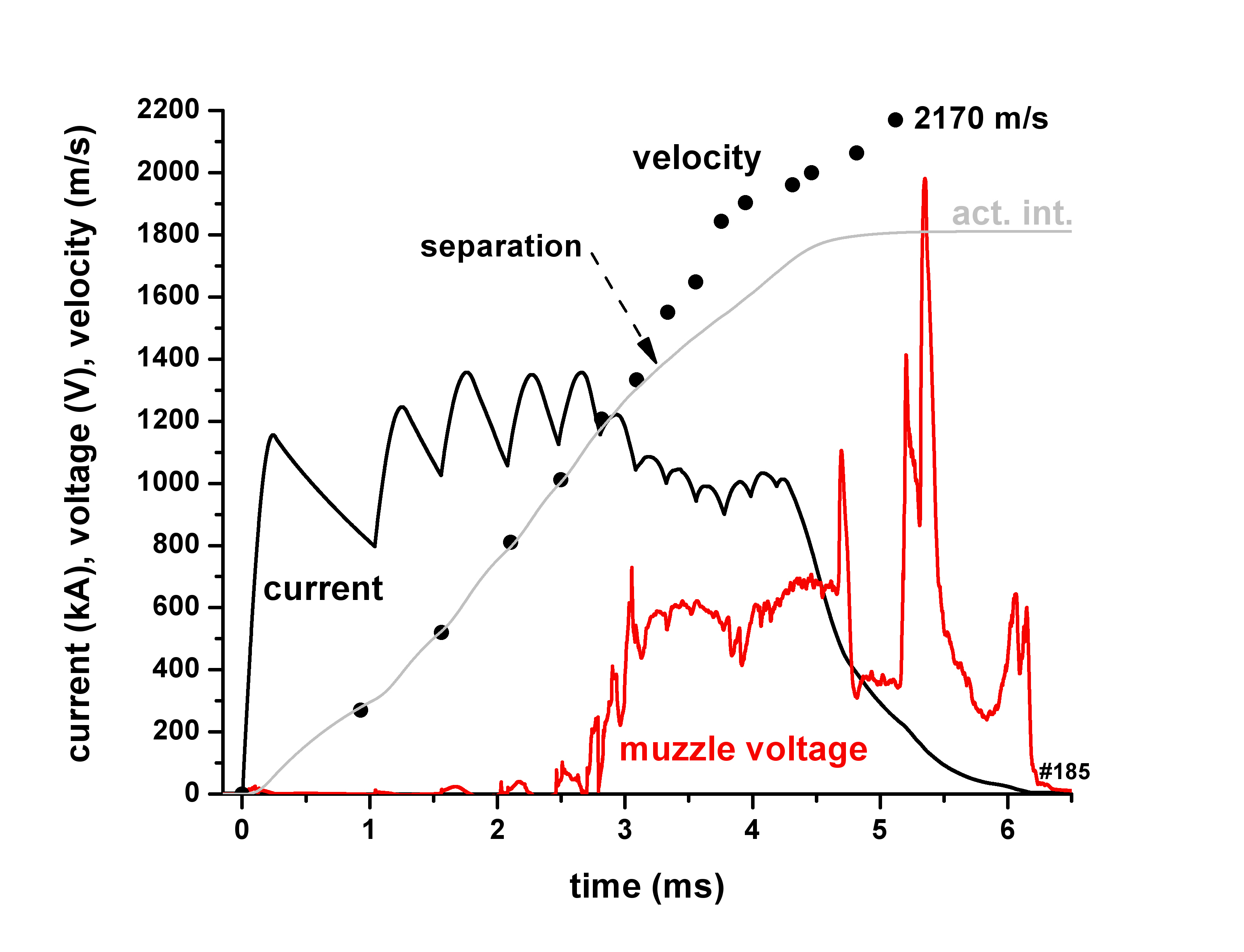}
\caption{Shot no.\,185.}
\label{t185}
\end{figure}

\begin{figure}[tb!]
\centering
\includegraphics[width=1.75in]{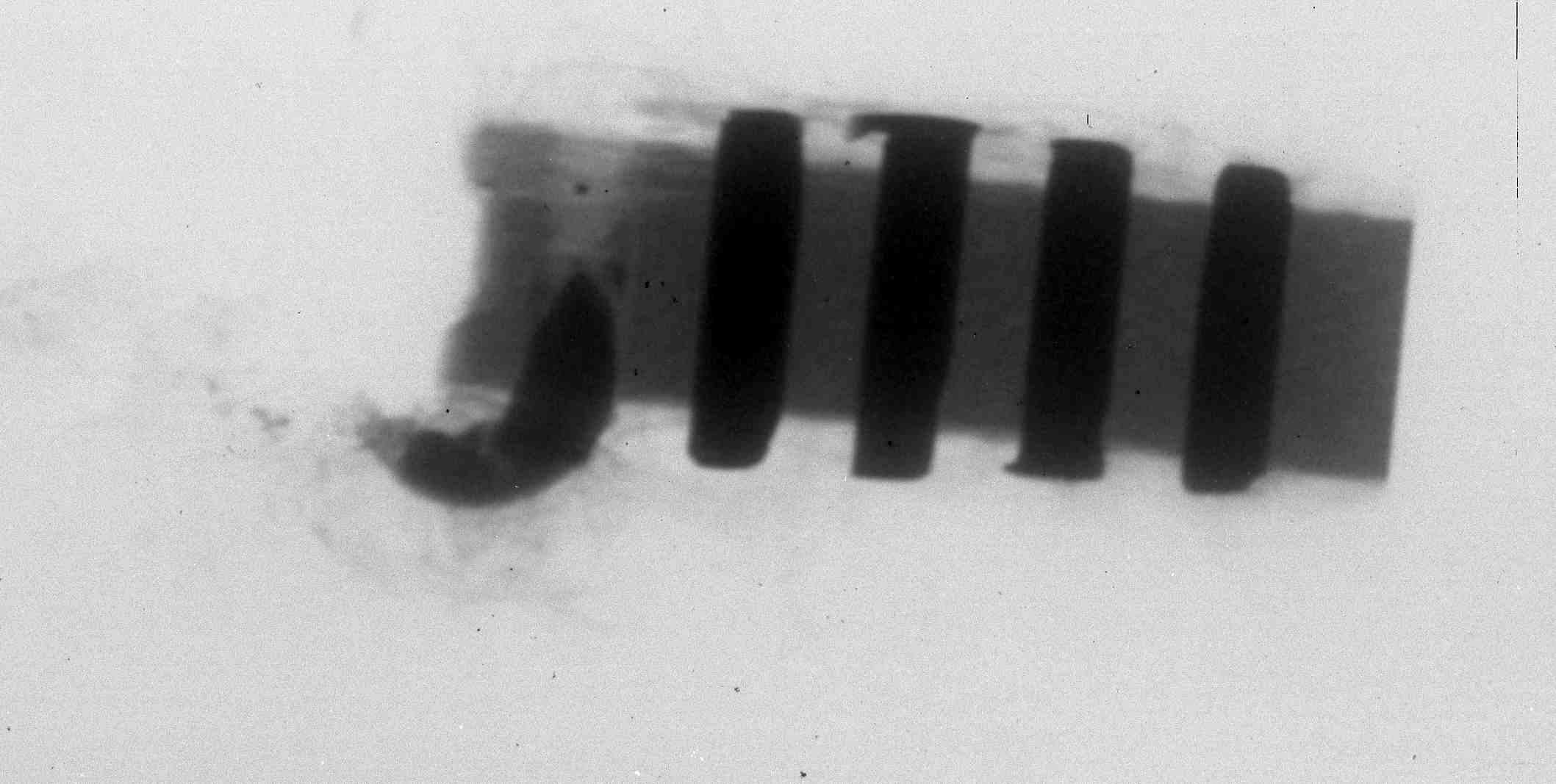}
\hfill
\includegraphics[width=1.6in]{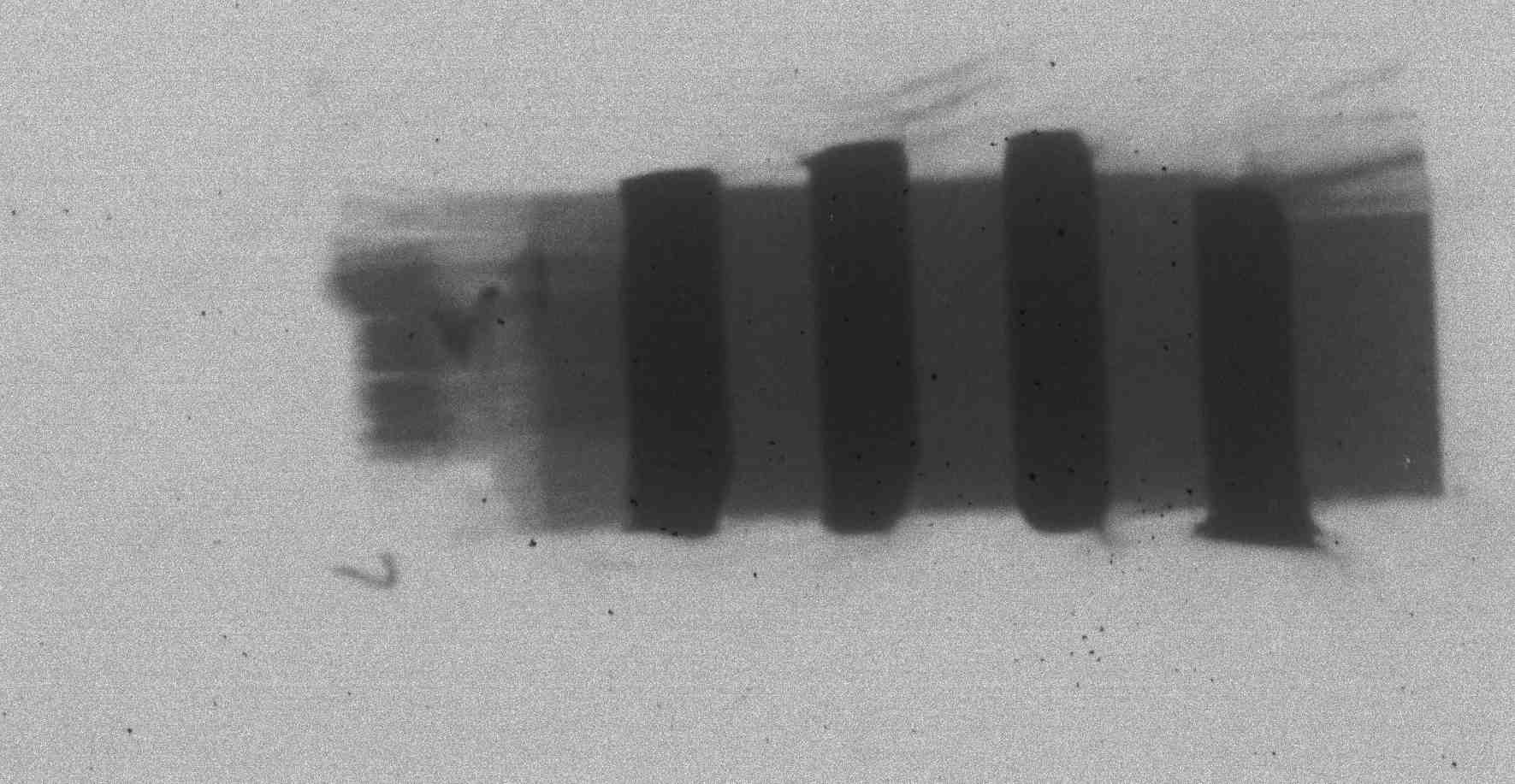}
\caption{X-ray picture of the payload part for shot no.\,184 (left)
and no.\,185 (right).}
\label{t184_t185_xray}
\end{figure}

After the introduction of the separating projectile, several
experiments with {\sc PEGASUS} were performed, to investigate the
behavior of this type of projectile. The primary
electrical energy was increased from 3.8\,MJ to 4.8\,MJ and the brush
diameter of the two booster brushes was changed from 8\,mm to finally
10\,mm. The key parameters for the shot series are itemized in
table \ref{shots_separating}.
\begin{table}
\centering
\begin{tabular}[tbh]{|l|c|c|c|}
\hline
\# & El. Energy & Velocity & Booster brush $\o$\\ \hline
181 & 3.8\,MJ & 1825\,m/s  & 8\,mm  \\ \hline
182 & 4.2\,MJ & 1880\,m/s  & 8\,mm  \\ \hline
183 & 4.2\,MJ & 1945\,m/s  & 9\,mm  \\ \hline
184 & 4.8\,MJ & 1820\,m/s  & 9\,mm  \\ \hline
185 & 4.8\,MJ & 2170\,m/s  & 10\,mm \\ \hline
\end{tabular}
\vspace{1ex}
\caption{Parameters of the shots performed with the separating projectile.}
\label{shots_separating}
\end{table}
\begin{figure}[tb!]
\centering
\includegraphics[width=.42\textwidth]{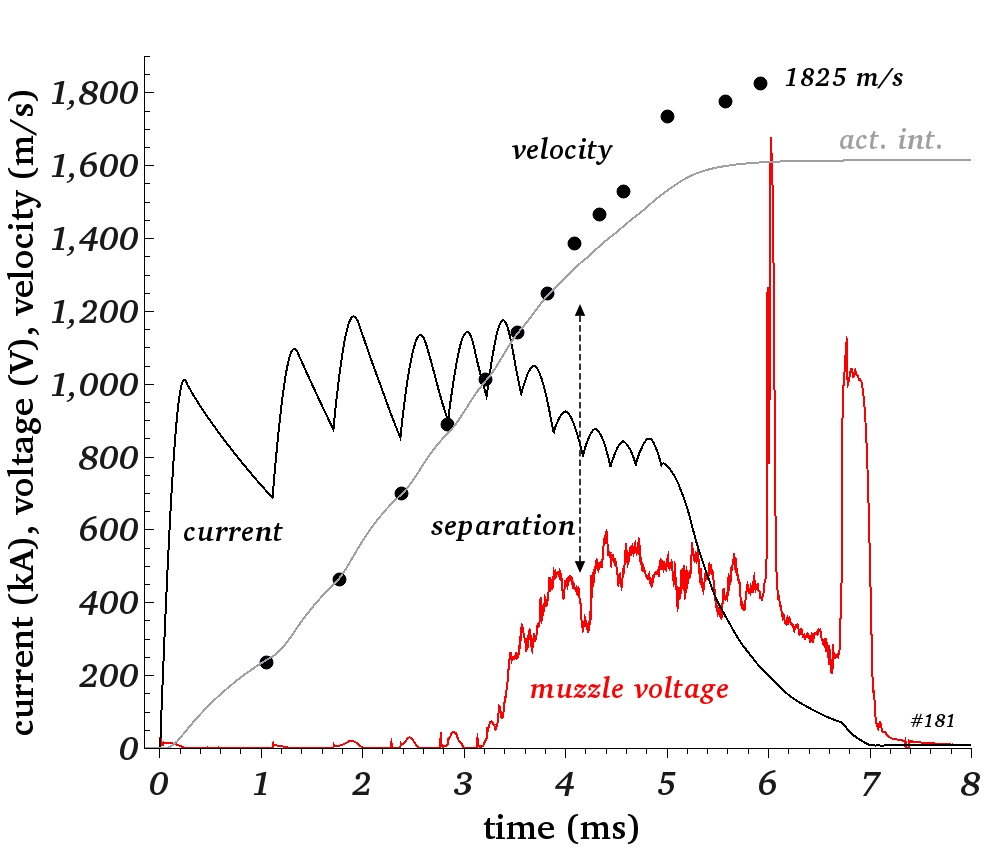}
\caption{Shot no.\,181.}
\label{t181}
\end{figure}
Figure \ref{t181} shows the current and the muzzle voltage
traces for the shot no.\,181. In addition to this, the velocity as
derived from measurements of the projectile passage using B-dot
sensors is shown using dots. The line label ``act. int.'' represents a
value that is proportional to the action integral ($\int \mbox{I}^2$dt). The
current reaches it's maximum value of 1.2\,MA at 1.8\,ms. The current
drop after 3.5\,ms is caused by the before mentioned capacity
reduction of 25\,\% for the banks no.\,8 to no.\,12. The muzzle voltage
distribution shows that the projectile experienced excellent sliding contact until
3.4\,ms, when transition occurred. The voltage rises quickly to a 
plateau of about 500\,V from 3.8\,ms until the shot out of the payload part at
5.8\,ms. The voltage trace of this shot looks different to the trace
from the standard projectile shot shown in figure \ref{t194_1}. The
inductive peaks and the linearly rising slope are not seen. The absence of the
inductive peaks indicates, that the current is spread out over a
larger area as compared to the standard projectile shot. 
The large rise of the muzzle voltage at 5.8\,ms witnesses the shot out of the
payload part of the projectile. Inspecting this peak closely, reveals
that the peak is actually a double peak. The earlier sub-peak can be
explained by current flowing through the copper cylinders (the
payload) in the front of the projectile. Later at 6.6\,ms a further
peak appears. At that time, the rear part of the projectile leaves the barrel.
The velocity of the projectile is measured, using b-dot probes
distributed along the barrel. Figure \ref{t181} shows that the
measured velocity follows the action integral until
approx. 4\,ms. After this time, the velocity rises stronger, than the action
integral. This can be understood as the point in time, when the front
part separates from the rear part, thus reducing the to be accelerated mass. 
For this shot, the velocity reached by the payload part is
1825\,m/s. The x-ray picture in figure
\ref{t181_t182_xray} (left)
\begin{figure}[tb!]
\centering
\includegraphics[width=.46\textwidth]{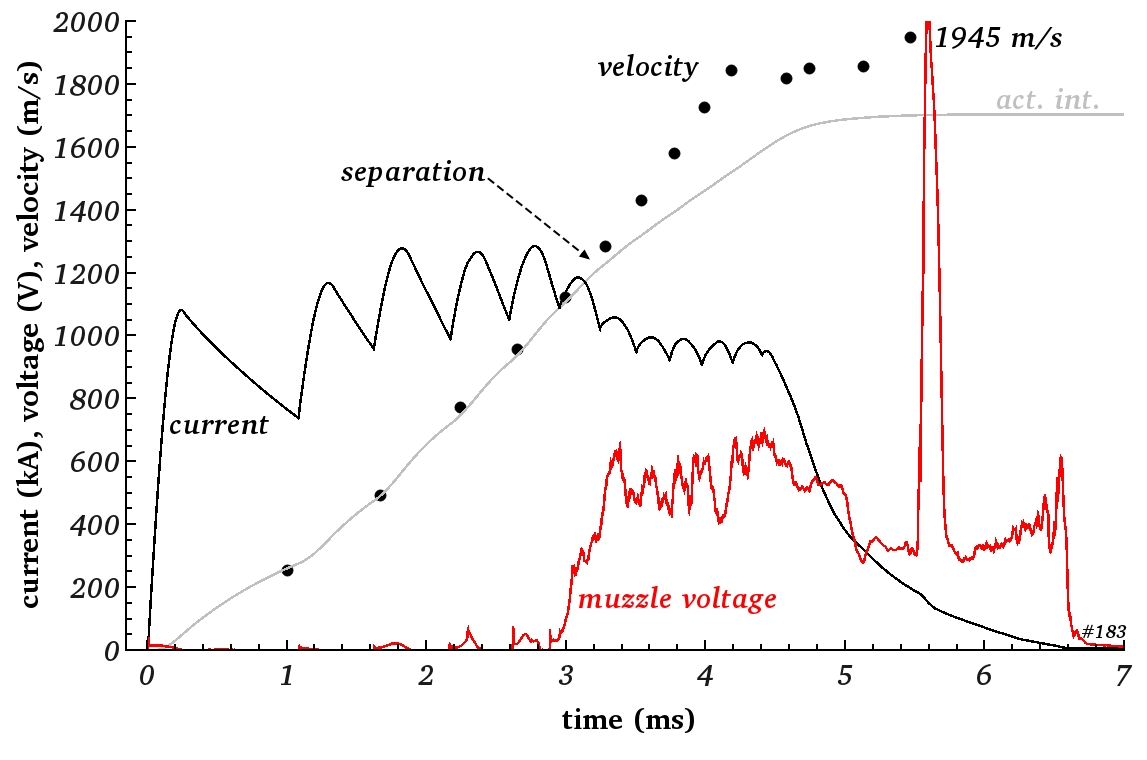}
\caption{Shot no.\,183.}
\label{t183}
\end{figure}
shows the payload part of the projectile during its free-flight
in the blue catch tank. The movement is from left to right. The
GRP body shows signs of delamination and the ends of the copper cylinders are
exposed. They show clear signs of current conduction during the shot,
thus supporting the discussion from above. 
In the next experiment, shot no.\,182, the primary energy was increased
to 4.2\,MJ. The measured end-velocity of 1880\,m/s was less than
expected. The free-flight x-ray picture
\ref{t181_t182_xray} (right) shows that the booster brush is nearly fully
eroded, indicating the development of a stronger plasma than in shot no.\,181. 
As a consequence of this data, the decision
was made, to increase the booster brush diameter to 9\,mm and redo the
experiment. The result of acceleration of this modified projectile is
shown in figure \ref{t183}.
The muzzle voltage graph shows about the same behavior as in shot
no.\,181. After transition at 3\,ms, a plateau of approx. 500\,V is
reached. Muzzle exit of the payload part is at about 5.5\,ms. The
measured velocity is 1945\,m/s. Inspecting the trace being proportional to the action 
integral and comparing it to the measured velocity of the projectile,
shows that the separation happened at approx. 3.3\,ms, just after the
current injection no.\,7. In shot no.\,181, the projectile separated
after injection no.\,9.   
\begin{figure}[tb!]
\centering
\includegraphics[width=.5\textwidth]{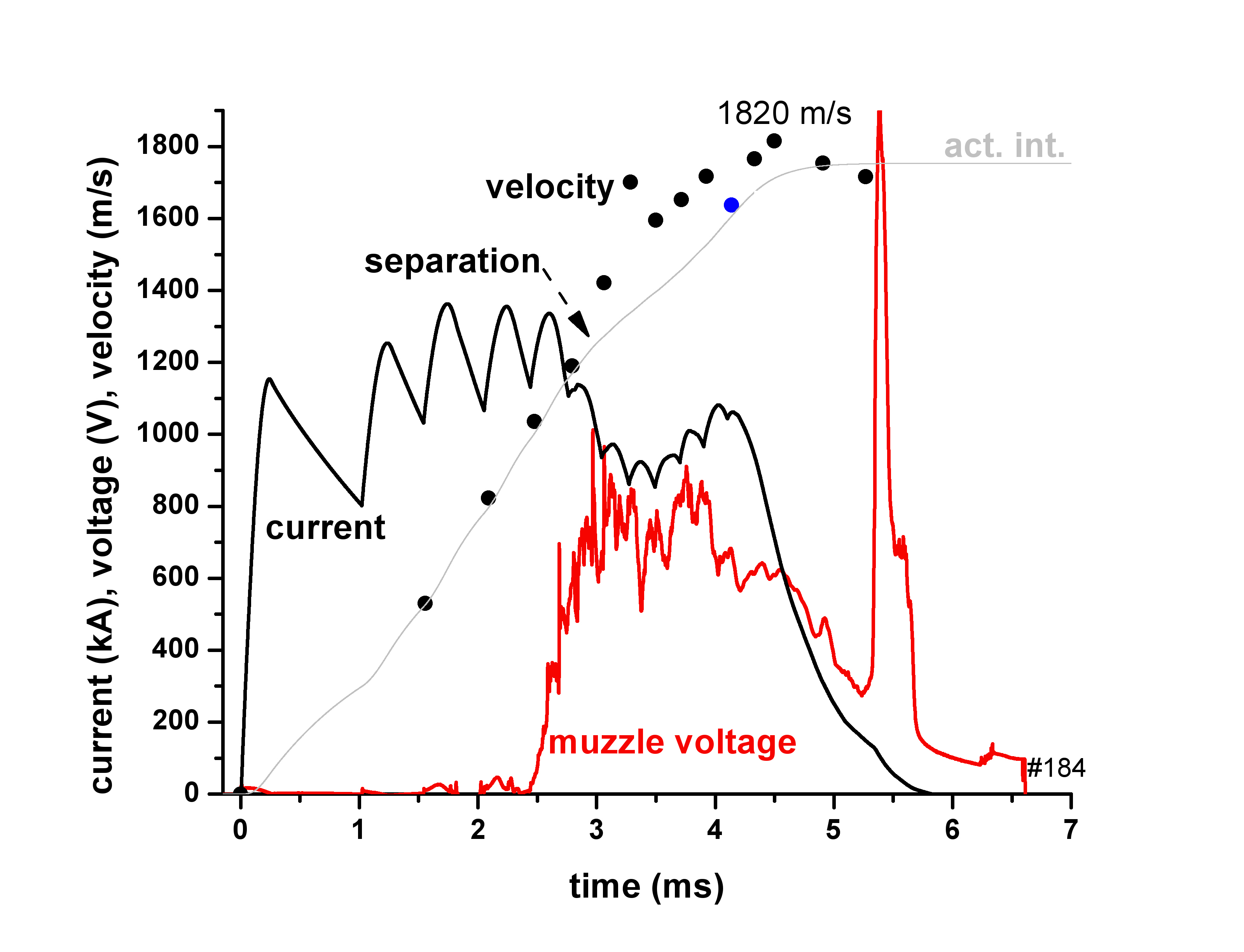}
\caption{Shot no.\,184.}
\label{t184}
\end{figure}
A further increase of the primary energy to 4.8\,MJ in shot no.\,184
resulted in the current and muzzle-voltage measurements shown in
figure \ref{t184}. A peak current of more than 1.35\,MA is reached,
but transition sets in early, at 2.5\,ms. In addition to this early
onset, the voltage rises up to more than 800\,V at about 3\,ms. This
behavior is different to the behavior seen in shots no.\,181 and
no.\,183. The maximum measured velocity is reached at 4.5\,ms, before
the projectile leaves the barrel. After this, the projectile
decelerates until shot-out. Figure \ref{t184_t185_xray} (left) shows a
strong material loss of the projectile body due to delamination.
The copper cylinders had contact with the rails and where most likely 
conducting current. It is visible that the booster brush is moving out
of its hole. Other x-ray pictures showing the top view, witness, that
the other brush is already lost before the x-ray picture is taken. The
movement and loss of the brush(es) is most likely the cause for the
development of the large values of the muzzle voltage and the deceleration of the
projectile in the final phase of the launch process. As the increase
of the booster brush diameter was successful for shot no.\,183, this
strategy was attempted again, and its diameter was increased to
10\,mm. Figure \ref{t185} shows the key parameter traces of this shot.
The current is again at about 1.35\,MA peak value. At 2.75\,ms transition sets
in, and until 3.2\,ms the trace resembles very much the standard
projectile behavior as shown in figure \ref{t194_1}. After 3.2\,ms an
approximate plateau of 600\,V to 620\,V is reached. The plateau is
followed by a peak at 4.8\,ms and shot out at 5.25\,ms. The split of
the shot-out peak indicates that there was current conduction in the
copper cylinders. The rear part of the projectile leaves the barrel at
about 5.9\,ms, producing a peak in the muzzle voltage trace. The measured
end-velocity for this shot is 2170\,m/s. The trace for the action
integral compared to the measured velocity shows, that separation occurred at
about 3.5\,ms. This is again after current injection no.\,7. The x-ray
picture in figure \ref{t184_t185_xray} (right) shows that the GRP holding
the booster brushes is broken during acceleration, thus indicating,
that the GRP holding the brushes is no longer able to cope with the forces 
during acceleration. For this reason a further increase in primary
energy with this type of projectile was not attempted.

\begin{figure}[tb!]
\centering
\includegraphics[width=3.5in]{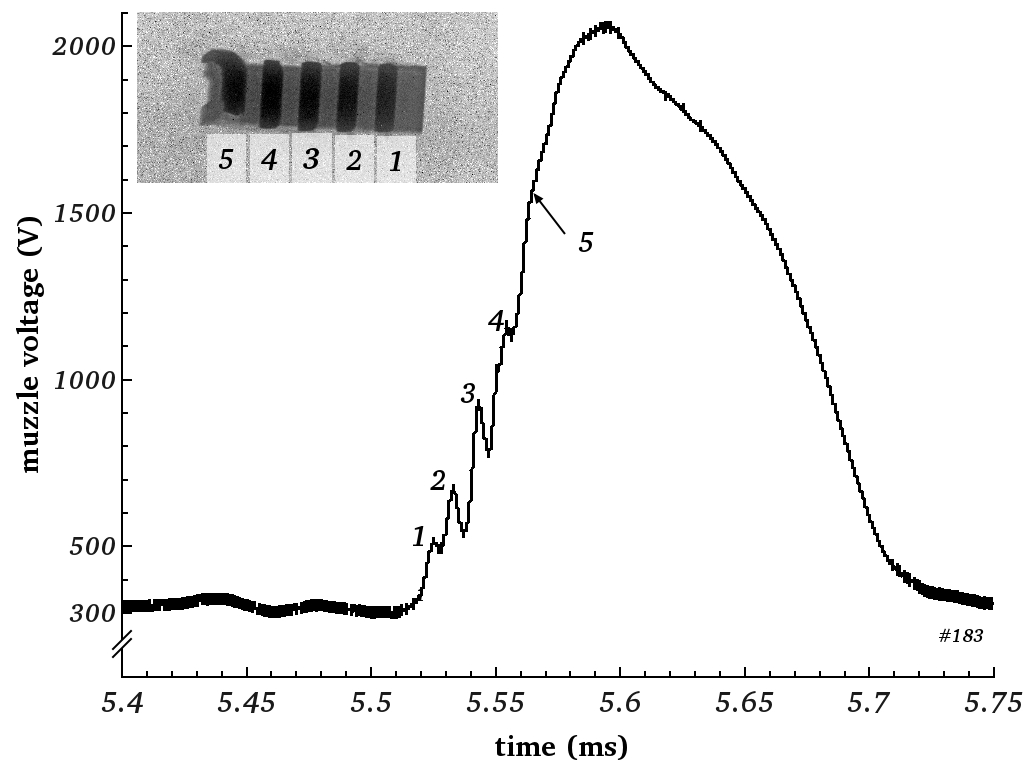}
\caption{Muzzle voltage for shot no.\,183 at shot-out. This is a
zoom-in of figure \ref{t183}. The inset shows the x-ray picture of the
projectile after it left the barrel.}
\label{t183_zoom}
\end{figure}

In figure \ref{t181} (shot no.\,181) the muzzle voltage peak caused by
the exit of the projectile from the barrel showed a sub-structure. This 
sub-structure was
interpreted as a sign for current conduction through the copper
cylinders. Support for this hypothesis comes from the x-ray pictures, that
show exposed copper cylinders with signs of mechanical wear from
sliding along the rails. In some of the figures one can also see the
copper evaporating from the hot copper cylinders. As an example, this
can be seen in figure \ref{t181_t182_xray} (left). For the projectiles
used here, the copper cylinders are 16\,mm apart (in direction of
flight). The same distance holds for the brushes, as well. When a
current carrying short circuit element like a brush or copper cylinder 
leaves the rails during shot-out, a plasma bridge will develop. This bridge 
connects the rail ends and the short circuit element. If another
short circuit element is in-between the rails, the plasma bridge will
eventually die out and the current will retract to this element. In
figure \ref{t183_zoom} the muzzle voltage trace from shot
no.\,183 enlarging the time during the shot-out of the payload-part of
the projectile is shown. Inspecting the raising slope from approx. 5.5\,ms on,
several small peaks on top of the slope are visible. The peaks are
labeled with number 1 to 5. In the upper left corner, the x-ray
picture of the projectile for shot no.\,183 is shown. The direction of
flight is from left to right. From right to left, the 4 rows of copper
cylinders (labeled 1 to 4) and the booster brush (labeled 5) are
visible. The muzzle voltage trace can be explained as follows. The
payload-part of the projectile approaches the muzzle while the current
drops rapidly (see figure \ref{t183}). At the time of shot-out the
current has approached a value of about 150\,kA. While most of the
current is flowing through the brush, some current is distributed
in-between the 8 copper cylinders. When the cylinders with row number
1 leave the rails, a plasma bridge is established and the resistance
increases. At the same the additional area bounded by the plasma
bridge is filled by the magnetic field, thus generating an inducting
effect. Due to this, the muzzle voltage raises. When the peak value
of peak number 1 is reached the bridge breaks and the current switches
back to the remaining current path still being in the barrel. During
the time the current switches to the rearward short circuit path the muzzle 
voltage drops. Then the subsequent rows disconnect one after the other
from the rails and the muzzle voltage peaks 2 to 4 are generated. When
finally the booster brush leaves the rails, the plasma bridge has to
carry the full remaining current. There is a slight change in slope,
when the brush fully has ceased to make contact to the rails, as then
the plasma bridge with a different resistance takes over.
This point is labeled with number 5 in the figure. After this time a
rising resistance due to the growing length of the plasma length
continues to generate a rising muzzle voltage. At one point in time the
current drops quicker than the resistance rises. For this shot this
happens at approx. 5.6\,ms. In table \ref{table_dist}, the distances
corresponding to the peak-to-peak time distance for a projectile with
a velocity of 1945\,m/s are calculated. It is seen that the calculated
values range between approx. 17\,mm and 21\,mm. The actual distance from
cylinder to the next cylinder (or brush) is 16\,mm. The hypothesis for
the slightly larger derived value is, that the contact is not 
a solid copper to copper contact, but instead an plasma connection
extending into the space around the cylinder.
Overall, the muzzle voltage behavior in combination with the x-ray
picture gives evidence, that for this shot, 
the copper cylinders took part in
establishing a short circuit route in-between the rails and were
carrying current.
\begin{table}
\centering
\begin{tabular}[tbh]{|l||c|c|c|c|c|}
\hline
peak number & 1 & 2 & 3 & 4 & 5 \\
\hline
time (ms)& 5.5245 & 5.5332 & 5.5435 & 5.5535 & 5.5642 \\
\hline
$\Delta$t to previous (x10$^{-5}$\,s)& \% &0.87 & 1.03 & 1.00 & 1.07\\
\hline
$\Delta$l to previous (mm)& \% & 16.9 & 20.0 & 19.5 & 20.8 \\
\hline
\end{tabular}
\vspace{1ex}
\caption{The time for the peaks labeled 1 to 5 in figure \ref{t183_zoom} 
is listed. For two following peaks, the time difference is calculated 
and converted to a distance by using the projectile velocity
of 1945\,m/s.}
\label{table_dist}
\end{table}

\begin{figure}[tb!]
\centering
\includegraphics[width=.48\textwidth]{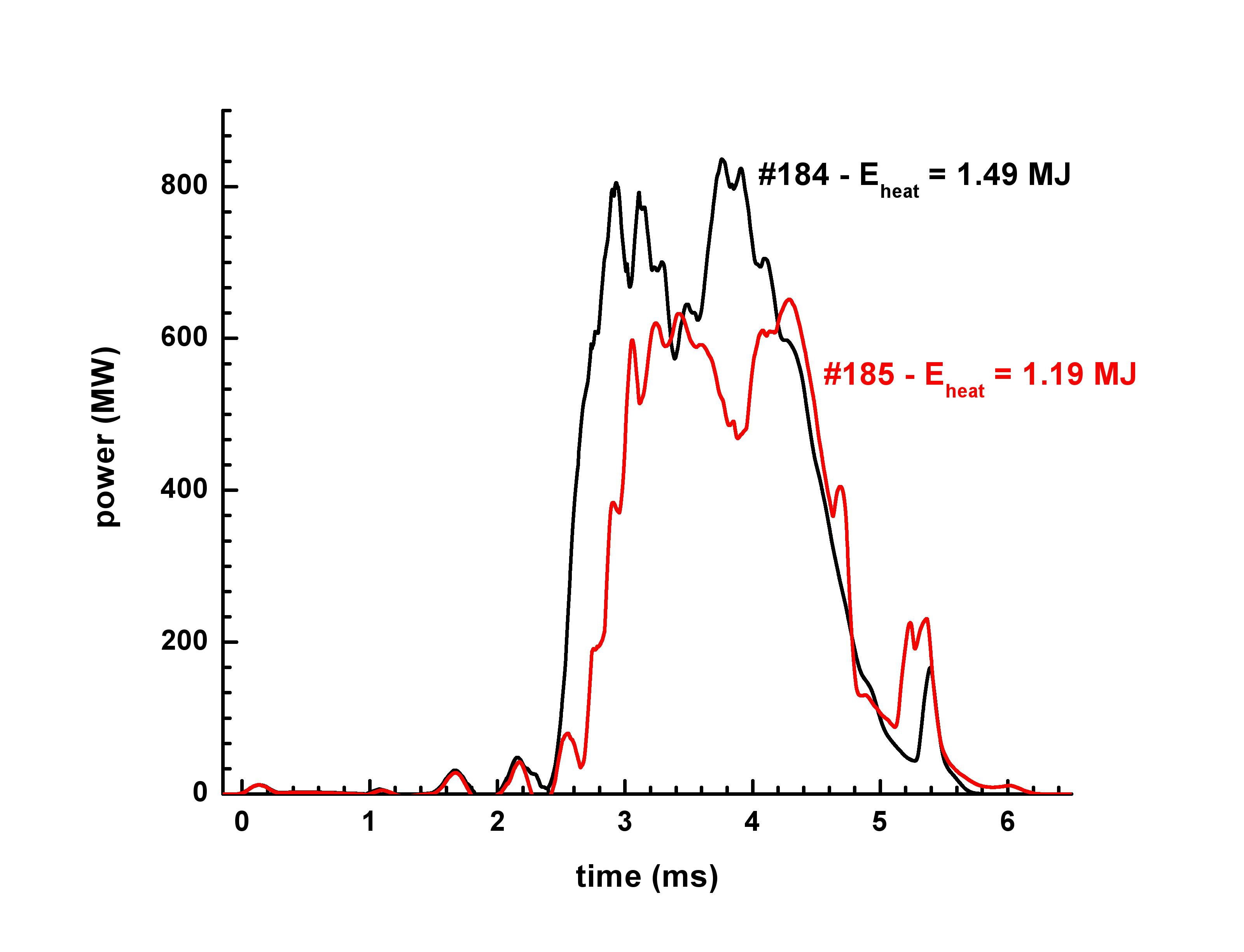}
\caption{Power across the rail-armature-rail contact for the two
4.8\,MJ shots (no.\,184 and no.\,185).}
\label{power}
\end{figure}

In table \ref{shots_separating}, 2 shots are listed with a primary
energy of 4.8\,MJ. The difference between these 2 shots is the
diameter of the brush. While the first shot (no.\,184) had a brush
diameter of 9\,mm, shot no.\,185 had this value increased to 10\,mm.
The kinetic energy of the payload part (approx. 400\,g) was 663\,kJ
and 942\,kJ, respectively. Shot no.\,185 has a payload part
kinetic energy that is 279\,kJ larger. In figure \ref{power} the power
acting on the rail-armature-rail interface is drawn. To suppress short
time fluctuations as seen, for example, in the muzzle voltage trace,
the curve was smoothed. Until 2.5\,ms, the onset of transition for
shot no.\,184, both curves are identical. The electrical power 
converted to heat during the acceleration at the rail-armature-rail
contact element is very low, basically all the available electrical
power at the rails is converted into acceleration of the projectile. After
transition the heating power reaches values of up to 830\,MW for shot no.\,184
and 660\,MW for shot no.\,185. Integrating the power over time, results in
the amount of energy that is converted into heat. This is 1.49\,MJ and
1.19\,MJ for shot no.\,184 and shot no.\,185, respectively. The
difference is 300\,kJ, very close to the 279\,kJ difference in kinetic
energy. This means that the additional available energy in shot
no.\,185 was converted to 90\,\% into kinetic energy of the payload
part of the projectile.

\section{Summary and Conclusion}
During an investigation spanning over several years, different payload
projectile types were designed and tested. In an earlier report
\cite{hundertmark_1} the experiments with an monolithic projectile and
the development path to the separating projectile was described. Here experiments
with a separating projectile type were presented.  A series of 5 shots 
was made and for each shot the parameters, current, muzzle voltage, velocity and
action integral were recorded. An inspection of the muzzle voltage
traces of these shots revealed a very different behavior as compared to the muzzle
voltage of the standard projectile with the same number of copper
brushes. If the contact of the brushes to the rails is not disturbed by strong
erosion or loss of brushes, the voltage raises after transition quickly to a 
plateau, while for the standard projectile, the voltage continues to
increase. For shot no.\,183 an enlarging of
the shot-out muzzle voltage peak showed that the payload copper
cylinders do take part in the sharing of the current. In addition to
this, traces of molten copper smoke from the heat developed by 
current conduction and mechanical friction can be identified in 
some of the free-flight x-ray pictures, pointing to the same conclusion.
As the energy was increased, it was necessary to enlarge the brush
diameter from 8\,mm to 10\,mm. This indicates that there is a positive 
effect of a larger armature mass being available. One clear problem at
the current {\sc Pegasus} railgun
for experiments with velocities above 2000\,m/s is the delamination 
of the GRP projectile body seen in the x-ray photographs in the 
region where the payload is mounted and the inability of the GRP to 
adequately support the pushing booster brushes. As a consequence of
this, the next design of the payload projectile will make usage of 
mechanically tougher materials. The currently preferred candidate
material is aluminum.
\section*{Acknowledgment}
The authors would like to thank for the professional support from the
people in the workshop and from those working with the experiments in the experimental hall.
This research was supported by the French and German Ministries of Defense.


%

\IEEEtriggeratref{6}

%
\IEEEpeerreviewmaketitle

\end{document}